\documentclass[reprint,amsmath,amssymb,aps,superscriptaddress,fleqn]{revtex4-2}
\usepackage{amsmath}
\usepackage{hyperref}
\usepackage{graphicx}
\usepackage{adjustbox}
\usepackage{bm}
\usepackage[dvipsnames]{xcolor}

\begin{document}

\title{A global microscopic description of nucleon-nucleus scattering \\ with quantified uncertainties}
\author{T. R. Whitehead}
\affiliation{Cyclotron Institute, Texas A\&M University, College Station, Texas 77843, USA}
\affiliation{Department of Physics and Astronomy, Texas A\&M University, College Station, Texas 77843, USA}
\affiliation{Facility for Rare Isotope Beams, Michigan State University, MI 48824, USA}
\author{Y. Lim}
\address{Max-Planck-Institut f{\"u}r Kernphysik, Saupfercheckweg 1, 69117 Heidelberg, Germany}
\affiliation{Institut f{\"u}r Kernphysik, Technische Universit{\"a}t Darmstadt, 64289 Darmstadt, Germany}
\affiliation{ExtreMe Matter Institute EMMI, GSI Helmholtzzentrum f{\"u}r Schwerionenforschung GmbH, 64291 Darmstadt, Germany}
\affiliation{Department of Science Education, Ewha Womans University, Seoul 120-750, Korea}
\author{J. W. Holt}
\affiliation{Cyclotron Institute, Texas A\&M University, College Station, Texas 77843, USA}
\affiliation{Department of Physics and Astronomy, Texas A\&M University, College Station, Texas 77843, USA}

\begin{abstract}
We develop for the first time a microscopic global nucleon-nucleus optical potential with quantified uncertainties suitable for analyzing nuclear reaction experiments at next-generation rare-isotope beam facilities. Within the improved local density approximation and without any adjustable parameters, we begin by computing proton-nucleus and neutron-nucleus optical potentials from a set of five nuclear forces from chiral effective field theory for 1800 target nuclei in the mass range $12 \leq A \leq 242$ for energies between $0\,{\rm MeV} < E \lesssim 150 \,{\rm MeV}$. We then parameterize a global optical potential for each chiral force that depends smoothly on the projectile energy as well as the target nucleus mass number and isospin asymmetry. Uncertainty bands for elastic scattering observables are generated from a full covariance analysis of the parameters entering in the description of our global optical potential and benchmarked against existing experimental data for stable target nuclei. Since our approach is purely microscopic, we anticipate a similar quality of the model for nucleon scattering on unstable isotopes.
\end{abstract}

\maketitle

\emph{Introduction} - Nuclear physics is approaching an exciting new era in which rare isotope beam facilities, such as FRIB, RIBF, FAIR, and Spiral2, will explore previously inaccessible regions of the nuclear chart that are important for understanding the origin of the elements \cite{FRIB,horowitz19,Benoit18,Kasen2017} and the properties of neutron stars \cite{Horowitz18,Brown17,Drischler21}. Rare isotope beam experiments will produce a flood of new data whose interpretation and connection to nuclear structure will be guided by theoretical modeling. Of particular importance in the context of nuclear reaction studies is the nuclear optical model \cite{PhysRevLett.3.96,Perey,Jeukenne77lda,JEUKENNE1976}, where the complicated (and in most cases intractable) problem of solving the $N$-body Schr\"odinger equation for nucleon-nucleus scattering in terms of fundamental two- and three-body forces is simplified by assuming the projectile nucleon interacts with an average single-particle potential generated by the target nucleus. Global phenomenological optical potentials \cite{KD03,bauge01,VARNER91} are the workhorse for theoretical modeling of nuclear reactions but are currently tuned to limited experimental data near nuclear stability. The worldwide radioactive ion beam program requires next-generation global optical potentials informed by microscopic nuclear theory based on high-precision nuclear forces \cite{epelbaum09,MACHLEIDT11,Whitehead20,Idini19,Rotureau18,Vorabbi18} that are able to reach into unexplored regions of the nuclear chart and provide quantified uncertainty estimates for reaction observables \cite{Nunes19,Nunesletter19}.

In the late 1960s the first phenomenological nucleon-nucleus optical potentials were limited to isotopes with mass numbers $A>40$ and low scattering energies of $E\lesssim 50$\,MeV. Phenomenological and semi-microscopic optical potentials \cite{VARNER91,KD03,Becchetti69,WILMORE1964673,Weppner,RAPAPORT197915,LI201262,LI200843,Furumoto19,Ruirui16,JEUKENNE1976,DelarocheILDA,bauge01} have improved dramatically since then, and today the most widely used optical potential of Koning and Delaroche \cite{KD03} is suitable to describe scattering phenomena for stable nuclei with $24<A<209$ up to projectile energies of $E \simeq 200$\,MeV. However, the reliability of such phenomenological optical potentials in the description of reactions involving exotic isotopes remains an open question. Dependable reaction models for rare isotopes are crucial for simulating the late-time freeze-out phase of r-process nucleosynthesis, where photodissociation and radiative capture processes are out of equilibrium and neutron-capture rates play an enhanced role in determining the final abundance pattern of r-process elements \cite{MUMPOWER16}. Neutron capture cross sections on neutron-rich isotopes cannot be directly measured with existing experimental techniques, but considerable efforts \cite{Spyrou14,Spyrou16} are being made toward measuring gamma strength functions and nuclear level densities that enter into the Hauser-Feshbach theory for radiative neutron capture in stellar plasmas \cite{RAUSCHER}. Such calculations also require as input the neutron-nucleus optical potential, and especially its imaginary part at low scattering energies \cite{GORIELY}. 

In the present work we construct the first microscopic global nucleon-nucleus optical potential based on an analysis of 1800 isotopes in the framework of many-body perturbation theory with state-of-the-art nuclear interactions from chiral effective field theory (EFT).
Compared to phenomenological \cite{VARNER91,KD03,Weppner,LI201262} or semi-microscopic optical potentials \cite{bauge01,Furumoto19,Ruirui16} that are directly fitted to nuclear reaction data, purely microscopic calculations may have greater predictive power for reactions involving exotic isotopes. Constructing optical potentials via the nucleon self-energy in finite nuclei or nuclear matter from chiral EFT \cite{Holt13omp,Holt15omp,Egashira,Rotureau18,Rotureau17,Vorabbi18,Toyokawa15,DURANT18,Idini19,Whitehead19,Whitehead20} is a promising route of inquiry since chiral EFT features realistic nuclear interactions based on the symmetries of low-energy QCD and a systematic expansion of nuclear forces \cite{WEINBERG79,epelbaum09,MACHLEIDT11,reinert18} that provides a method of quantifying theoretical uncertainties. In the present work, we consider microscopic nuclear forces at next-to-next-to-leading order (N$^2$LO) and N$^3$LO in the chiral expansion and with different choices for the cutoff ($\Lambda$) of the regulator function that suppresses high momentum states \cite{coraggio14,Entem03,Coraggio07,Marji13,Coraggio13,Sammarruca18}. The low-energy constants of the potentials are fitted to nucleon-nucleon scattering phase shifts, deuteron properties, and the triton binding energy and lifetime \cite{coraggio14}. From a covariance analysis of the five global optical potential parameterizations, we build a statistical ensemble of optical potentials from which we estimate scattering observable uncertainties.

\begin{figure}
	\begin{center}
		\includegraphics[scale=0.25]{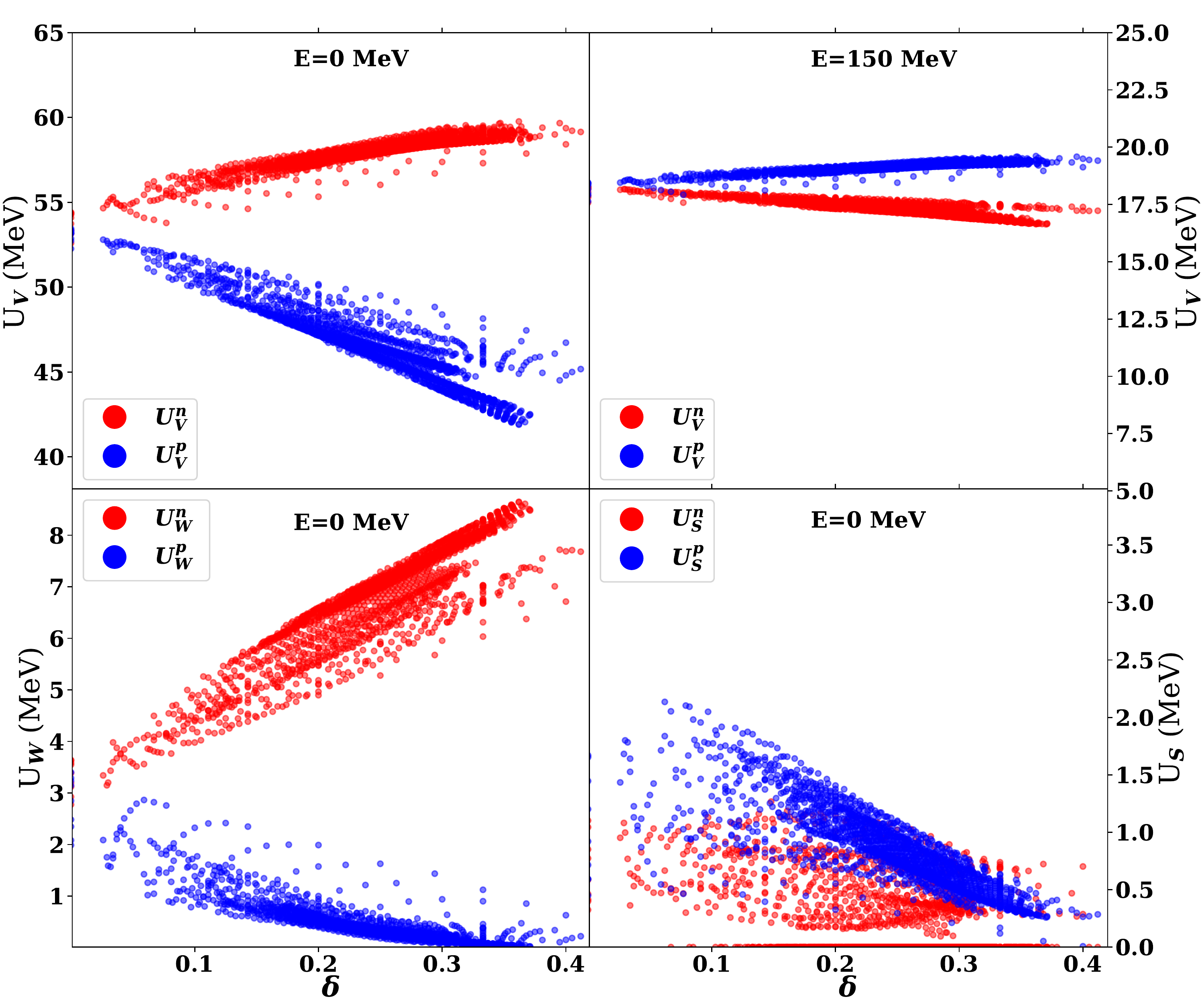}
		\caption{The left and right top plots show the depth of the real volume term at E = 0 MeV and E = 150 MeV for neutron and proton potentials as functions of the isospin asymmetry. The left and right bottom plots show the imaginary volume and surface depths respectively at E =0 MeV for neutron and proton potentials functions of the isospin asymmetry. The dots are values from the N$^3$LO, $\Lambda=450$ interaction for each of the target nuclei considered in this work.
			\label{depths}}
	\end{center}
\end{figure} 

\emph{Formalism} - We begin by calculating the nucleon self-energy $\Sigma(k,E(k))$ for $E>0$, which is equivalent \cite{PhysRevLett.3.96} to the optical potential, up to second order in many-body perturbation theory. Although no complete third-order calculation of the nucleon self energy in nuclear matter has been carried out to date, the sum of all third-order contributions to the equation of state have been shown \cite{Holt17} to be relatively small. Moreover, contributions to the nucleon self energy from resummed (particle-particle and hole-hole) ladder diagrams are on par with variations in the choice of nuclear potential \cite{Holt13omp,Rios20} and may be reduced through the inclusion of higher-order particle-hole diagrams \cite{Holt17}. The background medium is taken to be homogeneous nuclear matter with fixed density and isospin asymmetry in the thermodynamic limit. The calculation of the second-order diagrams involves intermediate-state propagators whose energies $E(k) = k^2/(2M) + \Sigma(k,E(k))$ are computed self consistently with the on-shell self-energy. In general the resulting self-energy is complex and energy dependent. In order to construct a nucleon-nucleus optical potential, we compute the nucleon self-energy over the range of densities and isospin asymmetries found in finite nuclei. Since the spin-orbit interaction vanishes in homogeneous nuclear matter, we employ the improved density matrix expansion \cite{bogner08,gebremariam10,Gebremariam10npa,KaiserHoltEDF} at the Hartree-Fock level to calculate the spin-orbit contribution to the nuclear energy density functional. In this formulation, the spin-orbit interaction is calculated at the Fermi energy and consequently does not have an explicit energy dependence. Density distributions for the target nuclei are calculated in mean field theory with Skyrme effective interactions \cite{Lim17} constrained by the same chiral interactions used to calculate the optical potential. In the present work, we neglect both nuclear deformation as well as time-odd mean fields when solving for the density distributions of odd-proton and odd-neutron nuclei. In particular, the inclusion of nuclear deformation (see e.g., Ref \cite{Furumoto19}) has been shown to improve the description of nucleon-nucleus scattering compared to experiment.

\begin{figure*}
	\begin{center}
		\includegraphics[scale=0.372]{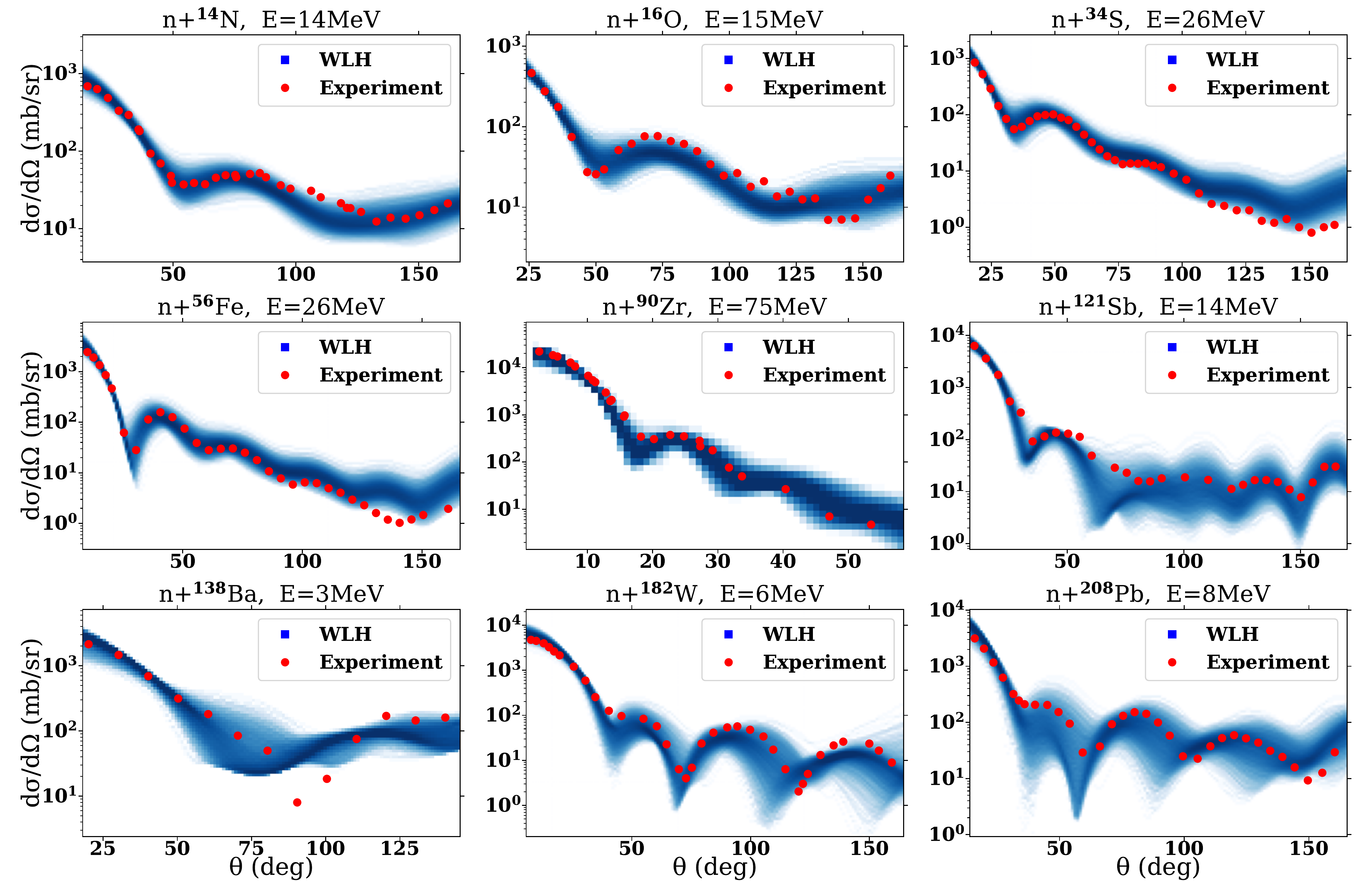}
		\caption{Neutron elastic scattering cross sections for a selection of target isotopes and varied energies. Results of the microscopic global optical potential constructed in this work are shown in shades of blue that represent cross sections calculated from 5000 random samples of the WLH optical potential. For a given scattering angle, the likelihood of a cross section value to be predicted by the WLH optical is represented by the color gradient where darker shades are more likely. Experimental data are shown as red dots \cite{Schmidt03,Glendinning1982,ALARCON86,Mellema83,Baba02,Rayburn59,BECKER66,ANNAND85,Roberts91}.
		\label{neutron_plot}}
	\end{center}
\end{figure*} 

The improved local density approximation (ILDA) is utilized to transition from a nuclear matter optical potential to that of a finite nucleus by folding the density- and isospin-asymmetry-dependent self-energy with the target nucleus density distribution $U_{LDA}(E;r)=U_{NM}(E;\rho (r),\delta (r))$ where $\rho=\rho_n+\rho_p$ and $\delta= (\rho_n-\rho_p)/ \rho$ . The ILDA is applied by integrating over the radial direction with a Gaussian form factor to account for the nonzero range of nuclear forces \cite{Jeukenne77lda,DelarocheILDA}:
\begin{equation}\label{eq:ilda}
U_{ILDA}(E;r)=\frac{1}{(t\sqrt{\pi})^3}\int U_{LDA}(E;r') e^{\frac{-|\vec{r}-\vec{r}'|^2}{t^2}} d^3r',
\end{equation}
where the range parameter $t$ represents the characteristic length scale of the interaction. The range parameter is derived in this work by calculating the root mean square radii of the local chiral NN interactions presented in Ref.\ \cite{Gezerlis14}. We use the average value of $t_C=1.22 \,{\rm fm}$ for the central terms of the optical potential and $t_{SO}=0.98 \,{\rm fm}$ for the spin-orbit term. In Refs.\ \cite{Whitehead19,Whitehead20} the effect of varying these range parameters was shown to be small.

\emph{Results} - Within the framework outlined above, we have developed in previous works \cite{Holt13omp,Holt15omp,Whitehead19,Whitehead20} proton and neutron optical potentials for stable calcium isotopes. In the present work we develop the first microscopic global optical potential that includes quantified theoretical uncertainties from nuclear forces, referred to as the Whitehead-Lim-Holt (WLH) global optical potential. The WLH global optical potential is built upon specific optical potentials for 1800 target nuclei with mass numbers $12\, < A < 242 \,$ and projectile energies $0\,{\rm MeV} < E < 150 \,{\rm MeV}$. The characteristic energy scale associated with nucleon-nucleus scattering is a combination of the projectile energy E and the kinetic energies of the target's constituent nucleons. Our maximum value of $E_{max} = 150 \,{\rm MeV}$ is heuristically estimated by identifying the projectile energy above which the theoretical uncertainties become uncontrolled. The set of target nuclei considered includes all stable and long lived isotopes, light and medium-mass bound isotopes out to the predicted neutron drip line of iron \cite{Stroberg21}, and heavier neutron-rich isotopes relevant to the r-process \cite{mumpower15}.

We fit the position- and energy-dependent optical potentials
\begin{eqnarray}
&&\hspace{.0in} U(r,E) = U_V(r,E) + i U_W(r,E) + i U_S(r,E) \label{phen} \\ \nonumber 
&&\hspace{.0in} + U_{SO}(r,E) \vec \ell \cdot \vec \sigma,
\end{eqnarray} 
to the commonly used Woods-Saxon form $f(r;r_i,a_i) = \frac{1}{1+e^{(r-A^{1/3}r_i)/a_i}}$ (for $U_V$ and $U_W$) and its derivative (for $U_S$ and $U_{SO}$). Functional forms for the $A$, $E$, and $\delta$ dependence of the Woods-Saxon geometry parameters and overall strengths were chosen in order to minimize the least squares fit while using as few parameters as possible. We used the following functional forms to define the global optical potential parameterization:
\begin{eqnarray}
{\cal U}_V &&= u_{V0}-u_{V1} E+u_{V2} E^2+u_{V3} E^3\\ \nonumber 
&&\pm (u_{V4} - u_{V5} E + u_{V6} E^2) \delta \\ \nonumber 
r_V&&=r_{V0}-r_{V1} E+r_{V2} E^2 - r_{V3} A^{-1/3} \\ \nonumber 
a_V&&=a_{V0} \mp a_{V1} E-a_{V2} E^2 -(a_{V3} - a_{V4} \delta) \delta
\label{vrealpara}
\end{eqnarray}
\begin{eqnarray}
{\cal U}_W&&= u_{W0}+u_{W1} E-u_{W2} E^2 + ( \pm u_{W3} -u_{W4} E )\delta \\ \nonumber 
r_W&&=r_{W0}+\frac{r_{W1}+r_{W2} A}{r_{W3} + A + r_{W4} E}+ r_{W5} E^2 \\ \nonumber 
a_W&&=a_{W0}-\frac{a_{W1} E}{-a_{W2}-E} + (a_{W3} - a_{W4} E)\delta \\ \nonumber
\label{wimpara}
\end{eqnarray}
\begin{eqnarray}
&&{\cal U}_S = u_{S0}-u_{S1} E- ( u_{S3}-u_{S4} E )\delta \\ \nonumber 
&&r_S=r_{S0}-r_{S1} E-r_{S2} A^{-1/3} \\ \nonumber 
&&a_S=a_{S0}
\label{simpara}
\end{eqnarray}
\begin{eqnarray}
&&{\cal U}_{SO} = u_{SO0}-u_{SO1} A \\ \nonumber 
&&r_{SO}=r_{SO0}-r_{SO1} A^{-1/3}  \\ \nonumber 
&&a_{SO}=a_{SO0}-a_{SO1} A,
\label{sopara}
\end{eqnarray}
\begin{figure*}
	\begin{center}
		\includegraphics[scale=0.372]{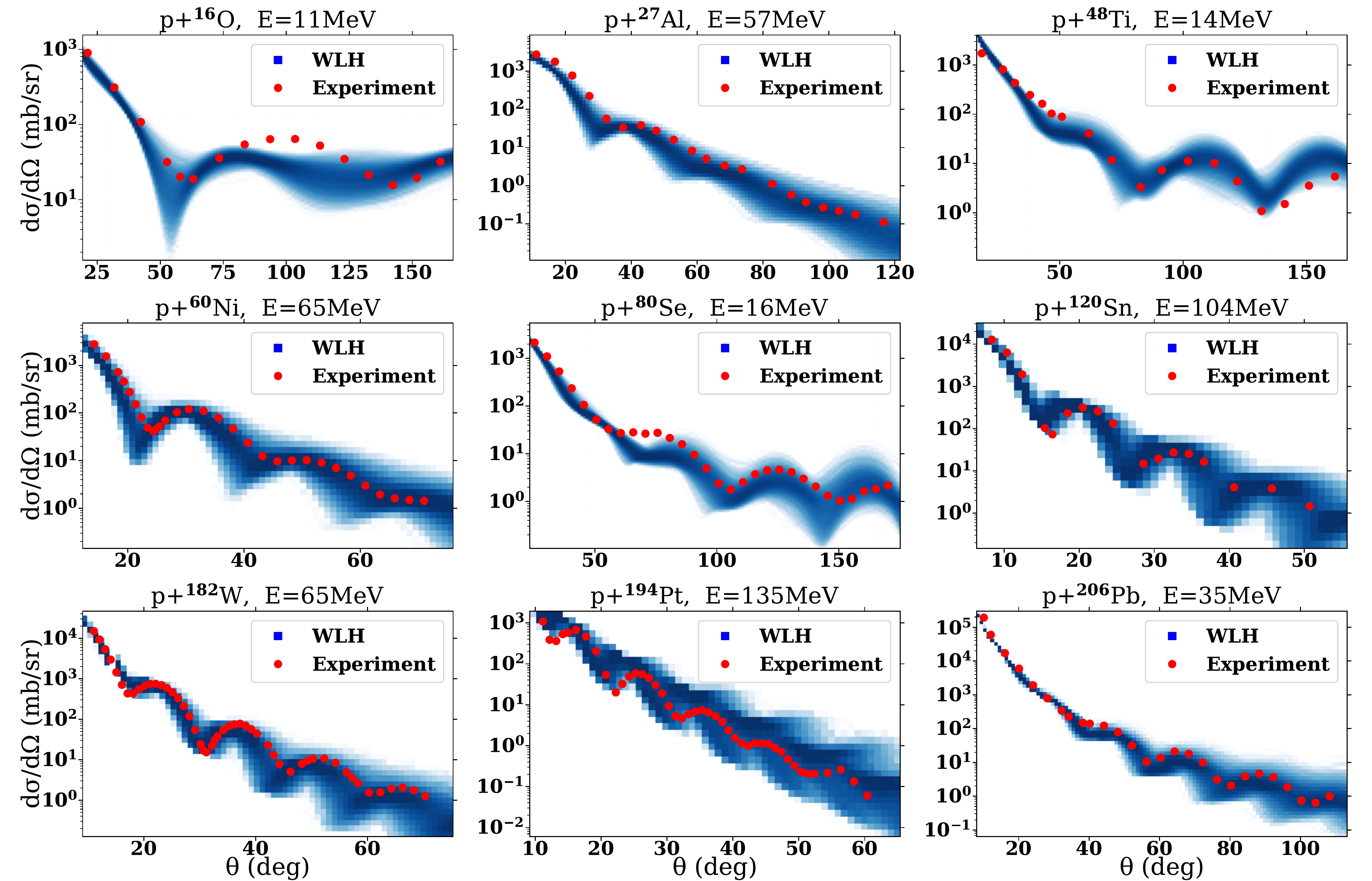}
		\caption{The same as Fig. 1, but for proton projectiles. Experimental data are shown as red dots \cite{Kobayashi60,Nonaka62,Kikuchi59,SAKAGUCHI81,DELAROCHE84,Kailas84,Ogawa86,SETHI90,FINCK83}.
			\label{proton_plot}}
	\end{center}
\end{figure*}
where the top signs in ($\pm , \mp$) refer to the value for proton projectiles and the bottom for neutron projectiles. The parameterization of the imaginary surface term is only valid for neutron projectile energies of $E \lesssim  40\,{\rm MeV}$ and proton projectile energies of $E \lesssim 20\,{\rm MeV}$. We find no imaginary surface peak beyond these energies.

To quantify the uncertainty coming from the choice of chiral potential, which we expect to be the dominant source of theoretical uncertainty, we begin by constructing global optical potentials based on each of the five chiral interactions used in the current work. The covariance matrix for all of the global optical potential parameters along with their mean values are used to generate a multivariate distribution which is sampled from to produce random parameter sets for the global optical potential. This enables one to estimate the uncertainty in a given reaction observable through many samples of the global optical potential. In particular, all results in the manuscript are generated from sets of 5000 samples.

\begin{figure*}
	\begin{center}
		\includegraphics[scale=0.372]{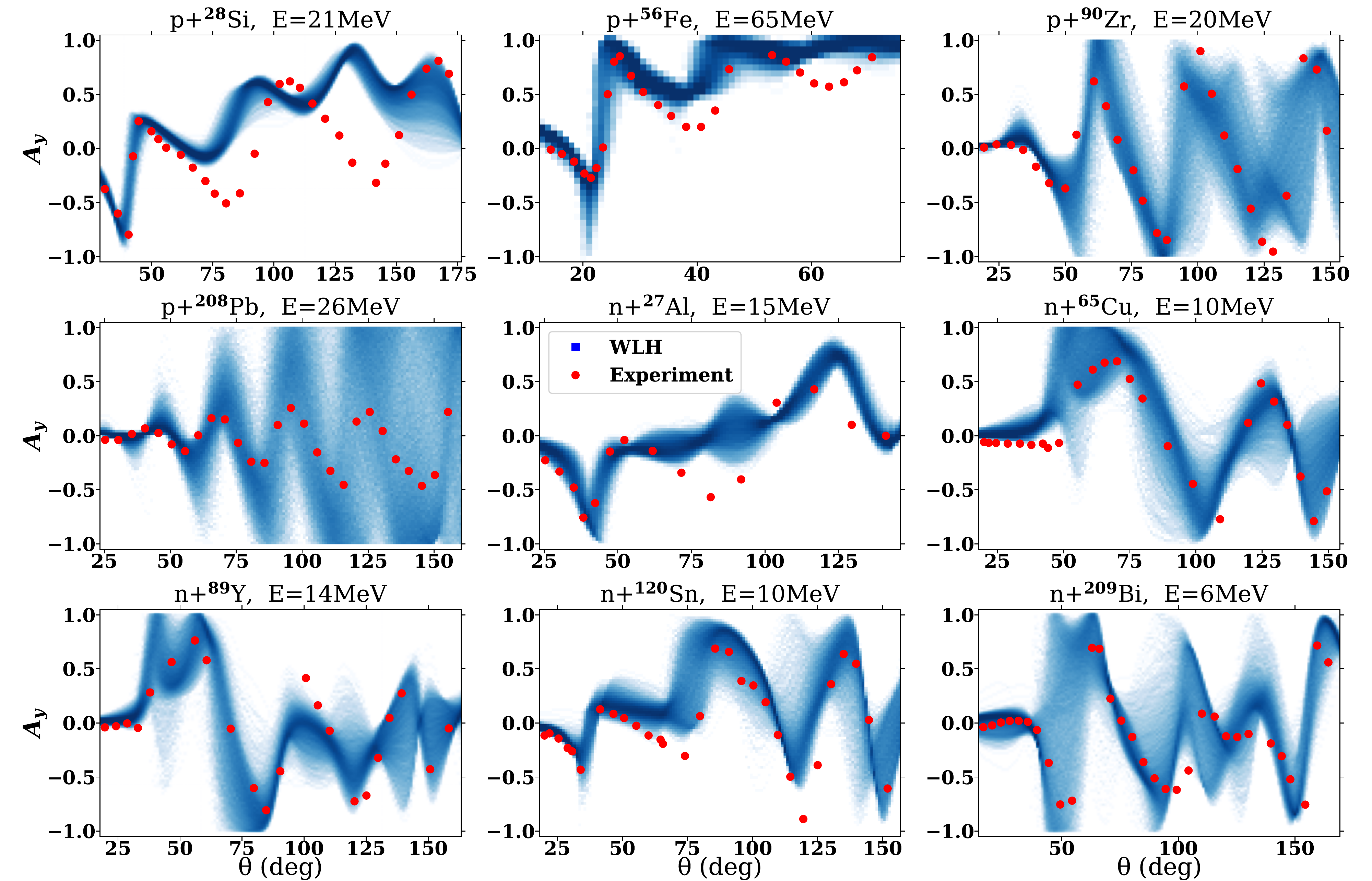}
		\caption{Similar to Fig. 1 and Fig. 2, but for the analyzing power. Experimental data are shown as red dots \cite{AOKI96,Leo96,Glashausser69,WATSON67,Nagadi03,Floyd83,Honore86,Guss89,Weisel96}.
			\label{ay_plot}}
	\end{center}
\end{figure*}

Next generation optical potentials for reactions involving exotic isotopes require realistic isovector terms that govern the behavior of the optical potential for asymmetric matter. In Fig. \ref{depths} we show the isospin asymmetry dependence of the real volume $U_V$, imaginary volume $U_W$, and imaginary surface $U_S$ depths using self-energies from the N$^3$LO, $\Lambda=450$ chiral interaction as a representative example. The top plots of Fig.\ \ref{depths} show that the real volume depth preserves the Lane form, $U = U_0 + \tau_z U_I \delta$, at both high and low energy. The real depth undergoes an isospin inversion where the isovector term changes sign near $E = 115 \,{\rm MeV}$, this can be seen by comparing the slopes for neutron and proton potentials in the top plots. The bottom plots show the imaginary volume and surface depths at $E = 0 \,{\rm MeV}$. For both terms there is an approximate linear dependence on the isospin asymmetry. The neutron imaginary terms shown in blue both decrease towards zero for large values of the isospin asymmetry. This is reasonable since the imaginary term vanishes at the Fermi energy, which approaches $E^n_F=0$ near the neutron drip line. The imaginary isovector term is of significant importance to the neutron capture rates on exotic isotopes involved in the astrophysical r-process \cite{GORIELY}.

To benchmark the WLH microscopic global optical potential, we calculate elastic scattering observables and compare to a wide range of experimental data for stable target isotopes. In Figs.\ \ref{neutron_plot} and \ref{proton_plot} we show neutron and proton differential elastic scattering cross sections for targets ranging from mass number A = 14 - 208 and projectile energies ranging from E = 3 - 135 MeV to demonstrate the performance of the WLH optical potential. In almost all cases, the experimental data lie within the probability contours associated with the cross sections predicted by the WLH global optical potential for both closed- and open-shell nuclei. In Fig.\ \ref{ay_plot}  we plot the analyzing power for neutron and proton projectiles and a variety of target nuclei. Again, the experimental cross sections typically lie within the microscopic uncertainty bands, which are however large for some target isotopes and projectile energies. The predictions from chiral effective field theory may be improved by the inclusion of second-order contributions to the nucleon spin-orbit potential \cite{KAISER10} and its explicit isospin-asymmetry dependence \cite{Kaiser12}. In future works, we also plan to implement N$^3$LO three-body forces \cite{tews13,drischler16,kaiser18,kaiser19,drischler19,holt19}.

\emph{Summary} - In the present work we have constructed the first microscopic global optical potential with quantified uncertainties. We suggest that the model may provide a foundation for analyzing and predicting nuclear reaction cross sections on target isotopes far from stability and for projectile energies up to E = 150 MeV. The global optical potential is expressed as a function composed of Woods-Saxon terms with parameters that vary smoothly in $E$, $A$, and $\delta$, which can be easily implemented into modern reaction theory codes. A Python script for sampling parameters of the WLH global optical potential may be found at \cite{WLH}. We show that experimental differential elastic scattering cross sections and analyzing powers are largely consistent within the uncertainties predicted by the WLH global optical potential, despite the fact that none of its parameters are fitted to nucleon-nucleus scattering data. Furthermore, we suggest that the WLH global optical potential can provide a prior distribution for Bayesian uncertainty quantification that incorporates empirical nucleon-nucleus data through appropriate likelihood functions. Such an approach may provide a powerful framework for constructing next-generation semi-phenomenological optical potentials for the theoretical modeling of nuclear reactions involving rare isotopes that are of interest to the nuclear reaction community moving into the rare isotope beam era in addition to topics that are central to nuclear astrophysics.

\begin{acknowledgments}
The authors thank H. Arellano for helpful comments on the manuscript. TRW thanks C. Drischler for insightful discussions. Work supported by the National Science Foundation under Grant No.\ PHY1652199 and by the U.S.\ Department of Energy National Nuclear Security Administration under Grant No.\ DE-NA0003841. Y. Lim was supported by the Max Planck Society and the Deutsche Forschungsgemeinschaft (DFG, German Research Foundation) – Project ID 279384907 – SFB 1245. Portions of this research were conducted with the advanced computing resources provided by Texas A\&M High Performance Research Computing and the Whitehead cluster.
\end{acknowledgments}

\end{document}